\documentclass[reprint,aps,prb,superscriptaddress]{revtex4-2}

\usepackage{amsmath,amssymb,graphicx,natbib,bm,array,physics,hyperref}
\usepackage{diagbox,longtable}
\usepackage[caption=false]{subfig}
\hypersetup{colorlinks = true,allcolors = blue }
\usepackage{bbm}
\usepackage{mathtools}
\usepackage{blkarray}
\usepackage{multirow}
\usepackage{outlines}

\DeclareMathOperator{\diag}{diag}

\newcommand{\z}{\mathbb{Z}}
\newcommand{\h}{\mathcal{H}}

\newcommand{\pos}{\Lambda}

\newcommand{\eref}[1]{(\ref{#1})}
\newcommand{\q}[1]{\mathrm{#1}}
\newcommand{\psup}[1]{P^{(#1)}}
\newcommand{\p}[2]{P_{#1}^{#2}}

\begin{document}
\title{Compact Wannier Functions in One Dimension}

\author{Pratik Sathe}
\email{psathe@lanl.gov}
\affiliation{Mani L. Bhaumik Institute for Theoretical Physics, Department of Physics and Astronomy, University of California Los Angeles, Los Angeles, CA 90095}
\affiliation{Theoretical Division, T-4, Los Alamos National Laboratory, Los Alamos, New Mexico 87545, USA}
\author{Rahul Roy}
\email{rroy@physics.ucla.edu}
\affiliation{Mani L. Bhaumik Institute for Theoretical Physics, Department of Physics and Astronomy, University of California Los Angeles, Los Angeles, CA 90095}

\date{\today}

\begin{abstract}
   Wannier functions have widespread utility in condensed matter physics and beyond. 
   Topological physics, on the other hand, has largely involved the related notion of compactly-supported Wannier-type functions, which arise naturally in flat bands. 
   In this paper, we establish a connection between these two notions, by finding the necessary and sufficient conditions under which compact Wannier functions exist in one dimension. 
   We present an exhaustive construction of models with compact Wannier functions and show that the Wannier functions are unique, and in general, distinct from the corresponding maximally-localized Wannier functions. 
\end{abstract}

\maketitle

\section{Introduction}
Wannier functions find applications in almost all areas of condensed matter physics.
In addition to providing a formal justification for the tight-binding approach~\cite{AshcroftSolid1976}, they help in understanding a broad range of properties of crystalline materials (see Ref.~\cite{MarzariMaximally2012} and the references therein). 

Maximally-localized Wannier functions are often used as a starting point in numerical studies~\cite{MarzariMaximally1997, Mostofiwannier902008}.
The localization properties of Wannier functions also have a direct bearing on the topological properties of electronic bands~\cite{ThoulessWannier1984,BezrukavnikovLocalization2018, MarcelliLocalization2022, Lu2021, SoluyanovWannier2011a}. 
Exponentially-localized Wannier functions can be constructed if and only if the Chern number and Hall conductivity of the corresponding filled band(s) are zero~\cite{ThoulessWannier1984,BrouderExponential2007, MonacoOptimal2018}.
An even stronger form of localization is compact support, a property of `compactly-supported Wannier-type functions', which feature in topological no-go theorems~\cite{Chen2014, Dubail2015,Read2017}. 

Compactly-supported Wannier-type functions necessarily exist in bands that are completely flat.
Flat-band Hamiltonians have attracted a lot of recent attention~\cite{LeykamArtificial2018}. 
They can host a variety of interesting phases, from the extensively studied integer and fractional quantum Hall effects in Landau levels~\cite{yoshioka2002quantum} to more recent examples such as 
unconventional superconductivity in bilayer graphene~\cite{KopninHightemperature2011, LuSuperconductors2019, CaoUnconventional2018, YankowitzTuning2019,MarchenkoExtremely2018,AokiTheoretical2020,VolovikGraphite2018}
and fractional Chern insulators~\cite{RoyFractional2011, BergholtzTopological2013, SpantonObservation2018}. 
Consequently, a substantial body of work focuses on systematically constructing flat-band models~\cite{MaimaitiCompact2017, MaimaitiUniversal2019a, Morales-Inostroza2016}, often exploiting compactly-supported Wannier-type functions, which are commonly also known as compact localized states in the context of flat bands.
When such wavefunctions form an orthogonal basis (a situation arising for example when all bands are flat~\cite{Ahmed2022b, Danieli2021, Danieli2021a, tovmasyan_preformed_2018}, or otherwise~\cite{KunoFlat2020}), the corresponding projector is Chern trivial regardless of lattice periodicity~\cite{Sathe2021, Kapustin2020}.

Similar to Wannier functions, compactly supported Wannier-type functions (when they can be constructed) span a band or a set of bands.
However, in general, Wannier-type functions are not mutually orthogonal and are therefore not truly Wannier functions (which are always mutually orthogonal).
In this work, we relate these two concepts and identify the necessary and sufficient conditions required for the existence of compactly supported Wannier functions (or compact Wannier functions in short) in one dimension, i.e. Wannier functions that vanish outside a finite region of the lattice.
Building on previous work~\cite{Sathe2021}, we show that compact Wannier functions (CWFs) can be constructed if and only if the band projector (or equivalently the appropriate single particle Green's function) is strictly local (SL).
Our proof leads us to an exhaustive construction in 1d of all possible models that have CWFs, and thus also of SL projectors.
We also show that CWFs are unique when they span a single band while CWFs that span multiple bands together are not unique. 
Finally, we relate compact Wannier functions to compactly supported Wannier-type functions by showing that the latter must be the same as the former when the former exist. 

Contrary to the intuition that compact localization is a stronger form of localization than exponential localization, we show that maximally-localized Wannier functions (MLWFs) are generally not compactly supported but are exponentially localized instead, even when CWFs can be constructed.
However, for nearest-neighbor-hopping projectors, we show that the MLWFs are compactly supported, and provide a construction to obtain such wavefunctions.
For systems without lattice translational invariance (LTI), we extend this construction to obtain generalized Wannier functions~\cite{KivelsonWannier1982a}.
Using this method, we show how to obtain an orthogonal basis of compactly-supported wavefunctions that spans the image of a non-LTI strictly local projector, with the size of each wavefunction being twice the maximum hopping range of the projector. 

This paper is organized as follows. 
In Sec.~\ref{sec:preliminary}, we present the notation used in this manuscript, followed by a discussion of the two families of wavefunctions that are the focus of this work: compact Wannier functions and compactly-supported Wannier-type functions.
Next, in Sec.~\ref{sec:CWFs}, we consider systems with LTI and establish the equivalence of CWFs and SL projectors in 1d systems and provide an explicit construction of CWFs for any given SL projector.
We also discuss the relationships between CWFs, MLWFs and CWTs for bands that have SL projectors.
In Sec.~\ref{sec:nn_proj}, we present some stronger results for nearest-neighbor hopping projectors, and provide constructions of MLWFs and generalized Wannier functions for projectors with and without LTI respectively.
We end with concluding remarks in Sec.~\ref{sec:conclusions}.

\section{Preliminary Discussion} \label{sec:preliminary}
\subsection{Notation and Setup} \label{subsec:notation}
We consider 1d tight-binding models with basis vectors denoted by $\ket{x,i}$, where $x$ denotes an integer-valued position index and $i=1,\dotsc,n$ denotes the orbitals centered within a cell\footnote{The tight-binding orbitals which form the basis may themselves be extended in real space, i.e., they may not have a strictly finite spatial extent.}.

Single-particle operators are represented in this basis by matrices with rows and columns labeled by $x$ and $i$.
Without loss of generality, we set the lattice constant to $1$.

We will find it convenient to represent the Hilbert space $\h_{\q{total}}$ in two forms, (1) as a tensor product of the position space and the orbital space $\h$, and (2) as a direct sum of spaces at each cell as follows:
\begin{align}
    \h_{\q{total}} &= \pos \otimes \h, \label{eq:H_tensor}\\
    \h_{\q{total}} &= \bigoplus_{x\in \pos} \h^x. \label{eq:H_oplus}
\end{align}
Here, $\h$ denotes the $n$ dimensional orbital space, while $\h^x$ denotes the Hilbert space at cell $x$.
$\pos$ denotes the lattice, being equal to $\z$ for infinite-sized systems and $\z_L$ for finite systems of size $L$.

We call a wavefunction compactly supported if it can be expressed as a linear combination of a finite number of tight-binding orbitals, i.e., orbitals whose labels (i.e. $x$ values) are associated with some finite region of the lattice.
We refer to the length of that region as the size of the wavefunction.
While this definition suffices for systems that are infinite in size, for a finite-sized system we also require that the size of a compactly supported wavefunction be smaller than the system length.

We now discuss strictly local (SL) projection operators in some detail since they play an important role in this manuscript.
Associated with any subspace of the single-particle Hilbert space is an orthogonal projector that projects onto the subspace.
We recall that an orthogonal projector is an operator $P$ that satisfies $P^2=P=P^\dagger$.
We call $P$ strictly local if 
\begin{align}
    \bra{x_1,i} P \ket{x_2,j} =0 \quad \text{ whenever }|x_1-x_2| > R \label{eq:defn_of_SL_p}
\end{align}
for some finite integer $R$.
We call the smallest value of $R$ the maximum hopping range of the projector.
For finite systems, a projector is SL only if its maximum hopping range is smaller the system size\footnote{This notion of strict locality of a projector is tied to the tightbinding basis and may not translate to strict locality in a continuum representation.}.

Let us now consider the case where the Hamiltonian has lattice translational invariance (LTI).
It is convenient to use the representation \eqref{eq:H_tensor}.
We define a plane-wave wavefunction $\ket{k} \coloneqq \frac{1}{\sqrt{L}}\sum_{x} e^{-ikx} \ket{x}$ where $L$ denotes the system size.
Following Bloch's theorem, a Hamiltonian eigenstate $\ket{\Psi_m(k)}$ corresponding to a band $m$ may be decomposed as 
\begin{align}
    \ket{\Psi_m(k)} =\ket{k} \otimes \ket{\psi_m(k)} \label{eq:bloch_wavefunctions}
\end{align}
where $\ket{\psi_m(k)}$ is a (normalized) column vector with $n$ entries.
We restrict $k$ to the first Brillouin zone (1BZ), i.e. $k\in [-\pi,\pi)$.
$k$ assumes only discrete values for a finite system with periodic boundary conditions.
The projector $P$ onto band $m$ may then be written as 
\begin{align}
    P = \sum_{k\in \q{1BZ}} \ket{\Psi_m(k)} \bra{\Psi_m(k)} = \sum_{k\in \q{1BZ}} \ket{k}\bra{k} \otimes P(k),     \label{eq:projector_in_k_space}
\end{align}
where $P(k)= \ket{\psi_m(k)}\bra{\psi_m(k)}$ is an $n\times n$ matrix.
(The summation sign is assumed to incorporate an appropriate normalization factor.)
Clearly, $P(k)^2=P(k)=P(k)^\dagger$, so that $P(k)$ is also an orthogonal projector.
The projector onto multiple bands may be defined similarly, by additionally summing over the band index $m$.

Strict locality of a projector with LTI is equivalent to the condition that each matrix element of $P(k)$ must be a Laurent polynomial in $e^{ik}$ (i.e. a Laurent series in $e^{ik}$ with non-zero coefficients only for powers of $e^{ik}$ between $-R$ to $R$, if $R$ denotes the maximum hopping range of the projector).

\subsection{Wannier and Wannier-type functions} \label{sec:wannier_and_wannier_type}
We will now review the definitions and properties of Wannier functions as well as Wannier-type functions.

For a Hamiltonian with LTI, we recall that the set $\{ \ket{\Psi_m(k)} \mid k \in \q{1BZ} \}$ of Bloch wavefunctions corresponding to an isolated Bloch band labeled $m$ is an orthonormal basis spanning the band.
A Wannier basis spanning the band consists of wavefunctions $\ket{w}_r$ for $r\in \pos$ defined via
\begin{align}
    \ket{w}_r = \frac{1}{\sqrt{L}} \sum_{k \in \q{1BZ}} e^{ikr} \ket{\Psi_m(k)}. \label{eq:Wannier_function_defn}
\end{align}
Each wavefunction $\ket{w}_r$ is (in general exponentially) localized around cell $r$ even in a tight-binding representation.
Since \eqref{eq:Wannier_function_defn} corresponds to a unitary mapping from the set of Bloch wavefunctions, the set $\{ \ket{w}_r \mid r \in \pos \}$ of Wannier functions is also an orthonormal basis spanning the band.
Furthermore, the wavefunctions corresponding to different values of $r$ are related by lattice translations.

We further recall that Wannier functions corresponding to a band are not unique since one may obtain a different set of Wannier functions with the replacement $\ket{\Psi_m(k)} \rightarrow \ket{\Psi_m(k)} e^{i\theta(k)}$ in \eqref{eq:Wannier_function_defn} for some real function $\theta(k)$.
Different choices of the gauge $\theta(k)$ result in different degrees of localization of the Wannier functions.
Identifying the gauge that corresponds to be most localized Wannier functions is one way of obtaining maximally localized Wannier functions (MLWFs).
For 1d systems that are infinite in size, the MLWFs are eigenstates of the $P\hat x P$ operator, where $\hat x$ denotes the position operator and $P$ denotes the projector onto the band of interest~\cite{MarzariMaximally2012}.
We will find this property useful while discussing MLWFs for bands with compact Wannier functions.

The discussion so far concerned single isolated bands.
In the case of a set of multiple bands (which may possibly be mutually intersecting) that are isolated from the rest of the bands, it is useful to construct `composite Wannier functions' that span the set of bands~\cite{Mostofiwannier902008}.
Specifically, one seeks to construct $q$ flavors of composite Wannier functions that together span a set of $q$ bands, without requiring that each flavor spans an individual band from the set of bands.
In this manuscript, we refer to composite Wannier functions also as simply Wannier functions.

Let us now discuss Wannier-type functions~\cite{Read2017}, which are in a certain sense a generalization of Wannier functions.
A flavor of Wannier-type functions consists of a localized wavefunction and all its lattice translates.
However, unlike Wannier functions, which are all mutually orthogonal, Wannier-type functions need not be linearly independent, let alone mutually orthogonal.
Thus, a set of $p$ bands may together be spanned by a set of $q$ flavors of Wannier-type functions with $q\geq p$.
Consequently, Wannier-type functions are in general not true Wannier functions.
We note that Wannier as well as Wannier-type functions are in general not Hamiltonian eigenstates, except in the case when the band is flat.
However, both varieties are eigenstates of the projector onto the band(s).

Since Wannier functions are also Wannier-type functions (with the additional condition of orthogonality), one can always construct Wannier-type functions for a band by virtue of the existence of Wannier functions.
Less trivially, it is also possible to construct \emph{non-orthogonal} Wannier-type functions for a band or a set of bands.
In the case of a single isolated band, for example, it is possible to obtain unnormalized but non-vanishing Bloch wavefunctions that span the band, which may then be used to construct Wannier-type functions. 
For instance, consider $\ket{\widetilde \Psi_m(k)} \coloneqq (2+ \sin^2 k) \ket{\Psi_m(k)}$ where $\ket{\Psi_m(k)}$ are the normalized Bloch wavefunctions from Eq.~\eqref{eq:bloch_wavefunctions}. 
Similar to the Wannier functions defined in Eq.~\eqref{eq:Wannier_function_defn}, we may analogously define a family of wavefunctions
\begin{align}
    \ket{\widetilde w}_r \coloneqq \frac{1}{\sqrt{L}} \sum_{k \in \q{1BZ}} e^{ikr} \ket{\widetilde \Psi_m(k)}. 
\end{align}
It is straightforward to see that $\{ \ket{\widetilde w}_r \mid r \in \pos\}$ is a set of wavefunctions that spans the band. 
While these wavefunctions are related to each other by lattice translations, they are not mutually orthogonal.
Thus, $\{ \ket{\widetilde w}_r \mid r \in \pos\}$ is a set of Wannier-type functions spanning the band.
One may similarly obtain Wannier-type functions that span multiple bands.

Interest in Wannier-type functions in recent years is primarily due to the fact that Wannier-type functions that are compactly supported always arise in Hamiltonians with flat bands.
Specifically, when a SL Hamiltonian has a flat band, the latter is necessarily spanned by such compactly supported Wannier-type functions or CWTs in short~\cite{Read2017}.
Furthermore, whenever CWTs span a band or a set of bands (regardless of band flatness), the band(s) have been shown to be topological trivial~\cite{Read2017}.
In the context of flat bands, CWTs are also known as compact localized states or CLSs in short and have been used extensively to construct model Hamiltonians that have flat bands~\cite{LeykamArtificial2018}.

While one may always construct Wannier and Wannier-type functions for a band or a set of bands, it is not always possible to construct such wavefunctions that are compactly supported.
Furthermore, in general, the existence of CWTs does not guarantee the existence of compact Wannier functions (CWFs).
(However, the existence of CWFs implies the existence of CWTs, since CWFs are orthogonal CWTs.)
For instance, while a flat band is always spanned by CWTs, in many cases, it is impossible to construct compact Wannier functions that span the band (see Ref.~\cite{Sathe2021} for some examples).
As we will show below, the condition that is necessary and sufficient for the existence of CWFs is that the corresponding projector is SL.

CWTs and CWFs cannot directly be defined for systems without LTI.
However, a natural generalization of Wannier functions to Hamiltonians without LTI is an orthonormal basis of localized wavefunctions that spans a subspace of the single-particle Hilbert space.
In one dimension, analogous to MLWFs, `generalized Wannier functions' have been defined to be the eigenstates of $P\hat x P$~\cite{KivelsonWannier1982a}.
A generalization of CWTs for Hamiltonains without LTI is a possibly non-orthogonal or even over-complete basis consisting of compactly supported wavefunctions that spans the occupied subspace of the single-particle Hilbert space.

\section{Compact Wannier functions} \label{sec:CWFs}
In this section, we consider systems with lattice translational invariance (LTI) and present various properties of CWFs in one dimension. 
We prove that CWFs can be constructed if and only if the band(s) they span are associated with a SL projector.
Next, we discuss the uniqueness of such bases, followed by a discussion of their relation to CWTs and MLWFs.

Except for the discussion pertaining to MLWFs, all our conclusions and methods are applicable to infinite lattices as well as finite lattices with periodic boundary conditions.
We consider flat as well as dispersive bands.
Our method for constructing CWFs works for single isolated bands as well as a set of composite bands, i.e. a set of bands that may be mutually intersecting, but which are isolated from the rest of the spectrum via gaps.
In the latter case, the projector in $k$-space, i.e. $P(k)$ corresponding to the composite bands, has a rank greater than one. 
We then obtain the corresponding composite compact Wannier functions.
On the other hand, in the case of entangled bands, i.e. a band or a set of bands that intersect with some other band(s) from the spectrum, it is a priori unclear how to assign a subspace to the band(s) of interest.
However, when a projector $P(k)$ can be assigned to an entangled band using a disentangling procedure such as from Ref.~\cite{SouzaMaximally2001}, we can still obtain CWFs using our procedure if $P$ is a strictly local projector.

Let us also mention related prior work on CWFs.
It has been noted that CWFs can be constructed in various models, for example in the dice model~\cite{VidalAharonovBohm1998}, the sawtooth lattice~\cite{huberBoseCondensationFlat2010a}, the diamond lattice~\cite{VidalInteraction2000a}, the Creutz ladder~\cite{creutzEndStatesLadder1999} and the Su-Schrieffer-Heeger model~\cite{SuSoliton1980} at certain parameter values.
In general, one dimensional Hamiltonians that only have flat bands always feature CWFs~\cite{Danieli2021}.
In specific models where CWFs can be constructed, they have been utilized to understand various phenomena such as existence of bound pairs of electrons~\cite{tovmasyan_preformed_2018}, Hilbert space fragmentation~\cite{nicolau_flat_2023} and flat-band quantum scars~\cite{KunoFlat2020}.
However, in all these works, CWFs have been discussed on a case-by-case basis, because, as pointed out in Ref.~\cite{tovmasyan_preformed_2018}, ``It is not known in general under which conditions\dots compact Wannier functions exist for a band''. 
To our knowledge, the only other prior work that studied compact Wannier functions in generality is Ref.~\cite{Sathe2021} by the current authors.

In Ref.~\cite{Sathe2021}, it was shown by construction that strict locality of 1d projectors is equivalent to the existence of an orthogonal basis of compactly-supported wavefunctions.
For projectors with LTI, however, the method of construction of such wavefunctions did not result in true compact Wannier functions.
The set of wavefunctions obtained from this procedure has the property that it is invariant only under translations by integer multiples of $2R$ (instead of $1$ required for Wannier functions), where $R$ denotes the maximum hopping range of the projector.
In contrast, the construction we present below always results in compact Wannier functions with the full lattice translational symmetry of the original lattice.
Unlike in Ref.~\cite{Sathe2021}, where a Gram-Schmidt orthogonalization procedure was used, here, we exploit the special structure of SL projectors revealed using singular value decompositions of their constituent blocks.

\subsection{Equivalence to Strictly Local Projectors} \label{subsec:equiv_to_SL_proj}
We will now show that in LTI systems, strict locality of a projector is equivalent to the existence of CWFs corresponding to the band(s).
First, we note that if a set of CWFs spans a set of $s\geq 1$ bands, then the corresponding projector $P$ is SL. 
This follows from the fact that $P(k)=\sum_i^s \ket{\psi_i(k)} \bra{\psi_i(k)}$, where $\ket{\psi_1(k)},\dotsc, \ket{\psi_s(k)}$ are the Fourier representations of distinct CWF flavors.
Clearly, $P(k)$ has matrix elements that are all Laurent polynomials (i.e. Laurent series with finite number of terms) in $e^{ik}$.
The proof of the converse is more subtle as we will now see.

To that end, we consider an SL projector with maximum hopping range $R\geq 1$.
(The case $R=0$ is trivial, since $P$ can be diagonalized with an intra-cell rotation thus giving us CWFs.)
As noted above, in $k-$space, it can be represented by a projection matrix $P(k)$ with elements that are Laurent polynomials in $e^{ik}$ with degrees $\leq R$.
Our procedure is iterative, with each step involving a SL unitary rotation $U_r (k)$ that reduces the maximum hopping range of the projector by $1$. 

Let us briefly discuss our strategy before presenting the iterative step.
The main idea is that since $P(k)$ is Hermitian, we can construct a (suitable) unitary operator $U(k)$ that diagonalizes $P(k)$ so that $U(k)^\dagger P(k) U(k) = \diag (1,\dotsc,1,0,\dotsc,0)$.
The inverse Fourier transforms of the columns of $U(k)$ that correspond to the eigenvalue $1$ are then Wannier functions that span the band(s) corresponding to $P(k)$.
Different valid choices of $U(k)$ correspond to different, equally valid constructions of Wannier functions.
We aim to construct a $U(k)$ with the special property of only having matrix elements that are Laurent polynomials in $e^{ik}$.

While $P(k)$ itself has matrix elements that are Laurent polynomials in $e^{ik}$ since $P$ is SL, it is not obvious that there exists a $U(k)$ that only has Laurent polynomial matrix elements.
To our knowledge, this property has not been proved in the literature.
Indeed, a generic valid $U(k)$ will not have Laurent polynomial matrix elements, unless carefully constructed.
To illustrate this point, consider the following SL projector:
\begin{align}
   P(k) = \frac{1}{6} 
   \begin{pmatrix}
  2 \sqrt{2} \cos (k)+3 &  -2 \sqrt{2} \sin (k)+i\\
 -2 \sqrt{2} \sin (k)-i & 3-2 \sqrt{2} \cos (k)
   \end{pmatrix} \label{eq:simple_example_projector}
\end{align}
Applying a standard procedure for diagonalizing a Hermitian matrix, we obtain a unitary matrix
\begin{align}U(k) &=
\frac{1}{\sqrt{6}}
    \begin{pmatrix}
 \frac{-2 \sqrt{2} \sin (k)+i}{\sqrt{3-2 \sqrt{2} \cos (k)}} & \frac{2 \sqrt{2} \sin (k)-i}{\sqrt{2 \sqrt{2} \cos (k)+3}} \\
 \sqrt{3-2 \sqrt{2} \cos (k)} & \sqrt{2 \sqrt{2} \cos (k)+3} \\
\end{pmatrix}
\end{align}
that diagonalizes $P(k)$, i.e. $U(k)^\dagger P(k) U(k) = \diag (1,0)$.
None of the matrix elements of $U(k)$ are Laurent polynomials in $e^{ik}$, and consequently, the Wannier functions obtained from $U(k)$ are exponentially localized instead of being compactly supported as desired.
Instead, our procedure guarantees the construction of a Laurent polynomial $U(k)$ which thus yields compact Wannier functions.
For the example above, our procedure results in the unitary matrix
\begin{align}
    U(k) =\frac{1}{\sqrt{6}}\begin{pmatrix}
        \sqrt{2} + e^{ik} & -i(\sqrt{2} - e^{-ik}) \\
        i(-\sqrt{2} + e^{ik}) & \sqrt{2} + e^{-ik}
    \end{pmatrix},
\end{align}
which diagonalizes $P(k)$ and only has Laurent polynomial entries.

We will now discuss the iterative step. 
Consider an intermediate step, at the start of which we have a projector $\psup r$ that has a maximum hopping range $r$ with $R\geq r\geq 1$. Note that $\psup r$ can be expressed as  
\begin{align}
    \psup r (k) = P_0 + \sum_{m=1}^r \left( P_m e^{ikm} + P_m^\dagger e^{-ikm} \right), \label{eq:P as sum of Pms}
\end{align}

First, we obtain a singular value decomposition (SVD) of $P_r$, which we conveniently write as 
\begin{align}
    P_r = \sum_{\sigma\neq 0} \sum_{j=1}^{n_\sigma} \sigma \ket{\psi_{\sigma, j}} \bra{\phi _{\sigma,j}}. \label{eq:Pr SVD}
\end{align}
Here, $\sigma$ denotes the non-zero singular values, and $n_\sigma$ denotes the degeneracy of $\sigma$. By construction, all the $\psi$'s are mutually orthogonal, and so are the $\phi$'s.

We note that $P_{r}^2 = 0$, which follows from $(\psup{r})^{2}=\psup{r}$, and \eref{eq:P as sum of Pms}.
Squaring \eref{eq:Pr SVD} and using $P_r^2 = 0$, we conclude that $\bra{\psi_{\sigma,i}}\ket{\phi_{\sigma',j}}=0$. Therefore, the set $S$ of all $\phi$'s and $\psi$'s together is a set of orthonormal wavefunctions. 
If $\vert S \vert < n$ (the number of orbitals per cell), then one can obtain enough wavefunction $\{\ket{\mu} \}$ such that together with $S$, they form an orthogonal basis for the orbital space $\h$.

We now implement a unit cell redefinition, so that all the $\phi$ wavefunctions at cell $x-1$, and all the $\psi$ and $\mu$ wavefunctions at cell $x$ together are grouped into a new cell at $x$.
This corresponds to a unitary transformation $U_r$, given by
\begin{align}
        U_r(k) = \sum_{\phi} e^{-ik} \ket{\phi} \bra{\phi} + \sum_{\psi} \ket{\psi} \bra{\psi} 
             + \sum_{\mu} \ket \mu \bra \mu, \label{eq:unit cell redefinition}
\end{align}
In this new basis, $\psup r$ transforms to $\psup {r-1} = U_r^\dagger \psup r U_r$ (which is also a projection matrix), and has a maximum hopping range of $r-1$ (for a proof, see Appendix~\ref{sec:Projector range reduction appendix}).

Thus, after at most $R$ iterative steps, the projector is an on-site hopping projector, so that one final $k$-independent rotation $U_0$ finally diagonalizes it. The total unitary transformation is 
\begin{align}
    P(k) &\rightarrow U^\dagger (k) P(k) U(k) = \diag(1,\dotsc,1,0,\dotsc,0) \nonumber \\
    \text{with }U(k) &= U_R(k) \dotsc U_1(k) U_0. \label{eq:full unitary transformation}
\end{align}
All matrix elements of $U(k)$ are Laurent polynomials in $e^{ik}$ by construction.
Hence, the inverse Fourier transforms of the columns of $U(k)$ corresponding to the $1$ eigenvalues are CWFs spanning the bands corresponding to $P(k)$.

We demonstrate this procedure by applying it step-by-step to an example SL projector in Appendix~\ref{sec:example}.

It is straightforward to show that the size of the CWFs can be at most $R+1$ cells, if the projector has a maximum hopping range $R$.
To that end, we note that each CWF obtained via the procedure described above is actually a bare orbital in the rotated basis determined by $U(k)$.
A CWF in the original orbital basis can be obtained by implementing the rotations in reverse on a bare orbital.
Since the application of each $U_i$ increases the size of a wavefunction in position space by $1$ cell, the CWFs cannot be larger than $R+1$ cells.
Additionally, at least one of the obtained CWF flavors must have a size of $R+1$ cells.
(Otherwise, the maximum hopping range of the projector can be inferred to be less than $R$, which contradicts the assumption.)

Therefore, the procedure for obtaining CWFs from an SL projector, implemented in reverse, is a recipe for an exhaustive construction of CWFs as well as of SL projectors with LTI. 
For example, if one wants to construct an SL projector spanning a single band in a three band model, and which has a maximum hopping range of $2$, then one can start with $\diag(1,0,0)$, and repeatedly transform it using intra-cell unitaries and unit-cell re-definitions. 

For an example of a non-trivial projector generated using this procedure, see Appendix~\ref{sec:example}.
In the absence of LTI, the crystal momentum $k$ is no longer a good quantum number, and as a result, this procedure is not directly applicable. 
Nonetheless, an orthogonal basis of compact wavefunctions can be constructed for an SL projector without LTI, as shown in Sec.~\ref{sec:supercell}.

\subsection{Relation to MLWFs} \label{subsec:relation_mlwfs}
Maximally localized Wannier functions (MLWFs) are defined as Wannier functions that have the least quadratic spread in space~\cite{MarzariMaximally1997, MarzariMaximally2012}.
In this subsection, we consider infinite sized 1d systems, in which case, MLWFs are eigenstates of the projected position operator, i.e. of $P\hat{x}P$~\cite{MarzariMaximally1997}.

In 1d, it has been shown that MLWFs have tails that die off exponentially (or faster)~\cite{MarzariMaximally1997}.
Since CWFs are compactly supported, it might seem reasonable to assume that they are also maximally localized.  
We will now show that this is not the case. 
Specifically, even when a set of bands is spanned by CWFs (or equivalently, the projector is SL), the corresponding MLWFs need not be compactly supported.

To that end, we provide a simple representative example. 
Consider a two band model with one band that is spanned by a flavor of CWFs, with the CWF at location $x$ being
\begin{align}
    \begin{split}
	\ket{\psi_x} = \frac{1}{\sqrt 6} (\sqrt{2} \ket{x,1} -\ket{x+1,1}+ \\ +\ket{x+1,2} + \sqrt{2}\ket{x+2,2}).
    \end{split}
\end{align}
Is it straightforward to express the $P\hat{x}P$ operator in terms of these wavefunctions. We find that
\begin{align}
\begin{split}
P\hat{x}P = \sum_x \Bigl( (x+1) \ket{\psi_x}\bra{\psi_x}  \\
 +\frac{1}{3\sqrt{2}} (\ket{\psi_{x+1}}\bra{\psi_x} + \ket{\psi_x}\bra{\psi_{x+1}})\Bigr).
 \end{split}
\end{align}
The MLWFs are then a set, consisting of a wavefunction $\ket{w_x}$ which is a simultaneous eigenstate of $P\hat{x}P$ and $P$, and all its lattice translates.
It is easy to see that no eigenstate of $P\hat{x}P$ (corresponding to a non-zero eigenvalue) can be compactly supported. 
(If that were the case, then $\ket{w_x}=\sum_{j=l}^{l+m} c_j \ket{\psi_j}$ for some $c_j$ and finite $l$ and $m>0$. 
We then find that $P\hat{x}P \ket{w_x}$ has a larger spatial spread that $\ket{w_x}$, leading to a contradiction.)

Thus for this example, MLWFs are not compactly supported, even though a CWF basis spans the band. 
However, the MLWFs are exponentially localized, which follows from Ref.~\cite{MarzariMaximally1997}.
We expect that the behavior seen in this example applies more generally; in other words, that the numerical construction of MLWFs will result not in CWFs, but in exponentially decaying Wannier functions.
We note that for finite systems with periodic boundary conditions, since the position operator is not well-defined, the MLWFs are not eigenstates of $P\hat x P$ and are instead obtained by maximizing an appropriate functional~\cite{silvestrelli_maximally_1999}. 
Nevertheless, even for these systems with periodic boundary conditions, one expects that the MLWFs will converge to the MLWFs of the corresponding infinite system as the system size is increased. Since the MLWFs of the infinite system do not match the CWFs considered here, the MLWFs for large enough finite systems are also not, in general, compactly supported.)

Even though CWFs are not always MLWFs, when a projector is nearest-neighbor hopping, the MLWFs are actually compactly supported. Furthermore, if such a projector does not have LTI, it is still possible to obtain `generalized Wannier functions'~\cite{KivelsonWannier1982a}, which are analogs of MLWFs. (See Sec.~\ref{sec:nn_proj} for proofs of both statements.)

\subsection{Uniqueness of Compact Wannier functions}
We will now show that if a single band is spanned by CWFs $\{ \ket{\psi_x} \vert  x\in \z\}$, then these CWFs form a unique set of Wannier functions (up to a phase) that are compactly supported for that band.

We prove this by contradiction. Suppose there exists another flavor of CWFs, $\{ \ket{\phi_x}\vert x \in \z \}$ which spans the band. Then every $\ket{\phi_x}$ can be expressed as a superposition of a finite number of $\ket{\psi_y}$'s, so that 
\begin{align}
\ket{\phi_x} = \sum_{k \leq j \leq l} c_j \ket{\psi_{x+j}},
\end{align}
with integers $k$ and $l$ such that both $c_k$ and $c_l \neq 0$.
Since the $\phi$'s are Wannier functions, they are orthogonal to their translates. With $T$ denoting translation by one unit cell, we thus have
\begin{align}
\begin{split}
0 &= \bra{\phi_x} T^{l-k} \ket{\phi_x}  \\
&= \sum_{k \leq j,m \leq l} c_j^* c_m \delta_{j+k, m+l} \\
&= c_l^* c_k,
\end{split}
\end{align} 
which is impossible since both $c_l$ and $c_k$ are non-zero. 
Thus, every other WF of the band is a superposition of an infinite number of translates of $\ket{\psi}$'s, implying that CWFs for a single band are unique up to inconsequential phases.

In contrast, the CWFs for multiple bands are not unique, so that when a set of multiple bands are together spanned by a set of composite CWFs, one can generate many such sets.
For example, if two band are together spanned by CWFs $\{\ket{\psi_x} \vert  x\in\z\} \cup \{ \ket{\phi_x}\vert x \in \z\}$, then $\{(\ket{\psi_x} + \ket{\phi_x})/\sqrt{2} \ \vert\ x\in\z\} \cup \{ (\ket{\psi_x} - \ket{\phi_x})/\sqrt{2} \ \vert \ x \in \z\}$ forms a distinct set of CWFs spanning the same set of bands.

\subsection{Uniqueness of Compact Wannier-type functions}
As discussed in Sec.~\ref{sec:wannier_and_wannier_type}, Wannier-type functions are a generalization of Wannier functions.
A flat band is always spanned by compactly supported Wannier-type functions (CWTs), which are called compact localized states or CLSs in that context~\cite{MaimaitiCompact2017} since they are also Hamiltonian eigenstates.

SL projectors can be associated with flat, i.e. dispersionless bands (for example, see Ref.~\cite{Ahmed2022b, Danieli2021, Danieli2021a}) as well as dispersive bands.
However, not all flat-band projectors are strictly local, and in general one cannot construct CWFs for flat bands.
Consequently, CLSs or CWTs are not, in general, expected to form an orthogonal basis.

We will now show that if a single flavor of CWTs spans a band described by an SL projector $P$, then the CWTs are actually the unique CWFs spanning the band. 
To that end, we will show that if a single flavor of CWTs spans a band, the CWTs form an orthogonal set of wavefunctions.
In Fourier space, the CWTs, say $\{ \ket{\psi_x} \vert x\in \z \}$ each of size $p$, correspond to a possibly unnormalized, but non-vanishing Bloch-like wavefunction $\ket{\psi(k)} = \sum_{j=0}^{p-1}\ket{\phi_j}e^{ikj}$ (with $\ket{\phi_0}\neq 0$ and $\ket{\phi_{p-1}}\neq 0$). 
Thus, the band projector $P(k)$ can be expressed as $P(k)=\frac{\mathcal{P}(k)}{\mathcal{Q}(k)}$, with $\mathcal{P}(k)=\ket{\psi(k)}\bra{\psi(k)}$ and $\mathcal{Q}=\bra{\psi(k)}\ket{\psi(k)}$. 
Each element of $\mathcal{P}$ and $\mathcal{Q}$ are then Laurent polynomials in $e^{ik}$ with degrees at most $p-1$.
Since each matrix element $\mathcal{P}_{ij}=\mathcal{Q} P_{ij}$, the degree of $\mathcal{Q}$ is $<p-1$ (or else, it equals $p-1$ and $P(k)$ has degree $0$, in which case the problem is trivial and hence we will not discuss it further).
Thus $\bra{\phi_{p-1}}\ket{\phi_0}=0$, with both vectors being non-zero.
One can thus iteratively implement unit-cell redefinitions similar to \eref{eq:unit cell redefinition} with the identification $\phi_0\rightarrow \phi$ and $\phi_{p-1}\rightarrow \psi$, and conclude that $U(k)\ket{\psi(k)} = \begin{pmatrix}
    1 & 0 \dotsc &0
\end{pmatrix}^T$, and therefore $\ket{\psi(k)}$ is normalized for all $k$ after all. 
In other words, the CWTs are actually CWFs, and it then follows from the uniqueness of CWFs that the CWTs are the unique CWFs.
We note that a similar procedure was used in Ref.~\cite{Danieli2021} for constructing compact localized states for Hamiltonians that only have flat bands.

\section{Nearest Neighbor Projectors} \label{sec:nn_proj}
Having discussed CWFs for SL projectors that have LTI, we will now provide some key results for projectors that are nearest-neighbor (NN) hopping in one dimension, with or without LTI.
Specifically, we will show that MLWFs in infinite systems are always compactly supported if the projector is NN. 
We will provide a procedure for such a construction.
In addition, we will show that even in the absence of LTI, one can construct analogs of MLWFs called generalized Wannier functions~\cite{KivelsonWannier1982a} for any NN projector.
We highlight that the MLWFs and generalized Wannier functions have a size of 1 or 2 cell only.
CWFs that have been obtained in the literature have primarily been of this type, but have been obtained in specific models instead of encompassing all possibilities.
(See, for example, the CWFs for Creutz ladder and the diamond chain, constructed in Ref.~\cite{tovmasyan_preformed_2018}.)

We consider infinite systems or finite systems with open boundary conditions in this section.
In either case and regardless of LTI, the wavefunctions we obtain are eigenstates of $P\hat{x}P$.
The procedure for constructing $P\hat{x}P$ eigenstates for NN projectors without LTI is also applicable to NN projectors with LTI. 
However, for the latter, we would like to obtain a set of Wannier functions, which have the property that the set they form is invariant under any lattice translation operation. 
Hence, the procedure developed for non-LTI systems requires modifications in order to obtain MLWFs for LTI NN projectors.
So, we first present the procedure for NN projectors without LTI, followed by that for NN projectors with LTI.
Finally, we discuss how to extend these results to SL projectors with larger hopping ranges.

\subsection{Nearest Neighbor Projectors without Lattice Translational Invariance}
\label{sec:NN no LTI}

First we describe some notation and conventions. 
We will say that $P$ `connects' two orbitals if $P$ has a non-zero matrix element between them. We find it convenient to define `hopping matrices' as follows. For any cell $x$, hopping matrices $P_i^x$ are defined as:
\begin{align}
    \p i x  \coloneqq \bra{x+i} P \ket{x}, \label{eq:hopping_matrix_def}
\end{align}
thus being matrices of size $n\times n$, where $n$ is the number of orbitals per cell. Since $P$ is NN hopping, for each $x$ only $\p{0}{x}$ and $\p{\pm 1}{x}$ can be non-zero. Furthermore, $\p{-1}{x+1} = (\p{1}{x})^\dagger$, since $P$ is Hermitian. 
We will also find it convenient to use the decomposition~\eqref{eq:H_oplus} for the total Hilbert space.

The primary tool in our procedure is again the singular value decomposition (SVD). 
Given an NN projector $P$, our objective is to obtain eigenvectors of $P$ that are also eigenstates of the $P\hat{x}P$ operator.
To that end, we leverage the properties of various blocks in the matrix representation of $P$.

Our procedure is iterative, with each step resulting in an intra-cell unitary rotation.
Each unitary rotation corresponds to a change of basis, and is accompanied by a reduction in the connectivity of the projector in the new basis. 
At the end of the procedure, we obtain states that are simultaneous eigenstates of $P$ as well as $P\hat{x}P$. 
We will now present this procedure.

The first step is to diagonalize $\p 0 x$ for each $x$. Thus, at each cell $x$, we obtain an orthogonal basis $\{\ket{x, \lambda_i }\}$ comprising eigenvectors of $\p 0 x$, that spans the local Hilbert space $\h^x$. 
Here, the index $i$ distinguishes degenerate states (if $\lambda$ is degenerate). 
For the steps that follow, we refer to orbitals as being the eigenvectors of the $P_0^x$ matrices.
When clear from the context, we will drop the position index and the degeneracy index if inessential.

We note that all eigenvalues of $\p 0 x$ matrices lie in $[0, 1]$. To see this, consider an eigenvector $\ket {\lambda}$ of $\p 0 x$ with eigenvalue $\lambda$. Then,
\begin{align}
    \p{0}{x} &= (\p{0}{x})^2 + (\p{1}{x})^\dagger \p{1}{x} +  \p{1}{x-1} (\p{1}{x-1})^\dagger \nonumber\\
    \implies \lambda - \lambda^2 &=  ||\p{1}{x} \ket {\lambda} ||^2 + ||\p{1}{x-1} \ket {\lambda} ||^2 \label{eq:lambda minus lambda square}\\
    \implies \lambda -\lambda^2 & \geq 0 \nonumber\\
    \therefore 0 & \leq \lambda \leq 1. \nonumber
\end{align}

Next, we note that $P$ connects orbitals at adjacent cells only if their eigenvalues add to $1$. In other words, $\bra{x+1, \lambda'} P \ket{x,\lambda} =0$, unless $\lambda + \lambda ' =1$. This follows from 
\begin{align}
    \begin{split}
    \p{1}{x} &= \p{1}{x} \p{0}{x}  + \p{0}{x+1} \p{1}{x}\\
    \implies \bra{\lambda'} \p{1}{x} \ket{\lambda} &= \bra{\lambda'}\p{1}{x} \p{0}{x} \ket{\lambda}  + \bra{\lambda'}\p{0}{x+1} \p{1}{x} \ket{\lambda} \\
    \therefore 0 &= (\lambda + \lambda' -1 ) \bra{\lambda'} \p{1}{x} \ket \lambda.
    \end{split}
\end{align}
A corollary is that if they exist, $\lambda=0$ states are annihilated by $P$, while $\lambda=1$ states are eigenvectors of $P$ with eigenvalue $1$. Additionally, it follows that if $\p 0 x$ has an eigenvalue $\lambda$, then at least one of $\p 0 {x\pm 1}$ must have an eigenvalue of $1-\lambda$. (Otherwise, we reach the contradiction that $P$ has eigenvalues other than $0$ and $1$.) 

With these properties at our disposal, we proceed to the next step, which is to obtain an SVD of all the $\p 1 x$ matrices simultaneously.
Specifically, we will show that judicious unitary rotations at each cell can bring all the $\p 1 x$ matrices into diagonal forms simultaneously.
We will show that such rotations can be obtained by implementing SVDs (with particular properties) of the $\p 1 x$ matrices sequentially.
[At face value, it is not obvious that this can be done, because (for example) an SVD of $\p 1 {x}$ can interfere with an SVD of $\p 1 {x-1}$, since the domain space and the target space respectively of the two are the same (i.e. $\h^x$).]
To explain how this can be done, we first discuss properties of the SVD of a single $\p 1 x$ matrix.

First, we note that orbitals corresponding to $\lambda=0$ and $\lambda = 1$ at $x$ (when they exist) can be ignored, since $\p 1 x$ annihilates them. 
For eigenvalues $\lambda\neq 0$ or $1$ of $\p 0 x$, as noted before, $P$ connects $\ket{x,\lambda_i}$ only to the $1-\lambda$ orbitals at cells $x\pm 1$.
Thus, $\p 1 x$ has a particular block structure, with non-zero blocks $\p {\lambda\rightarrow 1-\lambda} x$ that connect the $\lambda$ subspace at $x$ to the $(1-\lambda)$ subspace at $x+1$. 
If $\h_\beta^y$ denotes the eigenvalue $\beta$ subspace at cell $y$, then the block structure of $\p 1 x $ can be inferred to be:  
\begin{align}
    P_1^x = 
    \begin{blockarray}{c|c|c|c|cc}
    \h_{\lambda}^x & \dotsc & \h_{\mu}^x & \h_0^x & \h_1^x  \\
    \begin{block}{(c|c|c|c|c)c}
    \cline{1-5}
    P_{\lambda \rightarrow 1-\lambda}^x&0&0&0&0&\  \h_{1-\lambda}^{x+1} \\
    \cline{1-5}
    0&\ddots&0&0&0& \vdots \\
    \cline{1-5}
    0&0&P_{\mu \rightarrow 1-\mu}^x&0&0&\  \h_{1-\mu}^{x+1} \\
    \cline{1-5}
    0&0&0&0&0&  \h_1^{x+1}\\
    \cline{1-5}
    0&0&0&0& 0 &\h_0^{x+1}\\
    \end{block}
  \end{blockarray}, \label{eq:block_structure_P1}
\end{align}
where $\lambda,\dotsc,\mu$ denote the non-zero eigenvalues of $\p 0 x$.
[Note that it is possible for some eigenvalues to not be `paired' in the matrix representation above. 
For example $\p 0 x$ may have an eigenvalue $\gamma\in (0,1)$, but it is possible for $\p 0 {x+1}$ to not have an eigenvalue of $1-\gamma$. 
Similarly, it is possible that there may not be a $0$ or $1$ eigenspace at $x$ or $x+1$. In all such cases, the corresponding rows/columns should be understood as being absent in the block structure above.]

We exploit this block structure, and obtain an SVD of $\p 1 x$ by combining the SVDs of each of the blocks obtained independently. 
Thus, we can obtain unitary rotations within each $\lambda$ subspace at $x$ and each $1-\lambda$ subspace at $x+1$, so that (in the new basis) $\p 1 x$ connects every $\lambda$ orbital at $x$ with either zero, or exactly one `partner' orbital corresponding to eigenvalue $1-\lambda$ at cell $x+1$.
Henceforth, orbitals at $x$ and $x+1$ will refer to these post-rotation orbitals.

Having implemented these rotations, let us now discuss how an SVD of $\p 1 {x-1}$ can be obtained without disturbing the partner structure between cells $x$ and $x+1$.
To that end, we first note that whenever orbital $\ket{x,\lambda}$ has a partner orbital $\ket{x+1, 1-\lambda}$, $P$ does not connect $\ket{x,\lambda}$ with any orbital at cell $x-1$. (This follows from the fact that $P^2=P$ is NN-hopping, so that $\p{1}{x} \p{1}{x-1} = 0$.)
On the other hand, it also follows that if an orbital $\ket{x,\lambda}$ with $\lambda\neq 0,1$ does not have a partner orbital at cell $x+1$, then $\ket{x,\lambda}$ must be connected to at least one orbital (which must have eigenvalue $1-\lambda$) at cell $x-1$, and thus $\p 1 {x-1}$ must have a block $\p {1-\lambda,\lambda} {x-1}$ that connects these subspaces.

Thus, $\p 1 {x-1}$ maps the $1-\lambda$ orbitals at cell $x-1$ only onto those $\lambda$ orbitals at $x$ that do not have partner orbitals at $x+1$. 
Thus, an SVD of $\p{1-\lambda \rightarrow \lambda}{x-1}$ can be obtained that leaves untouched the $\lambda$ orbitals at cell $x$ that have partners at $x+1$. 
Such an SVD also ensures that all the $\lambda$ ($\neq 0,1$) orbitals at $x$ now get exactly one partner state, either at $x+1$ or at $x-1$.
[This along with \eref{eq:lambda minus lambda square} implies that the non-zero singluar values of $\p {\lambda,1-\lambda} x$ are all equal to $\sqrt{\lambda(1-\lambda)}$.]
Implementing such SVDs of all its blocks, we thus obtain an SVD for $\p 1 {x-1}$ which respects the SVD of $\p 1 x$.
In a similar fashion, one can obtain an SVD for $\p 1 {x+1}$ without disturbing the partner structure between cells $x-1$ and $x$.

The procedure ends when SVDs of all the $\p 1 {}$s are obtained sequentially following the prescription above. At the end of the procedure, the connectivity of the projector is greatly reduced in the rotated orbitals, and has the property that every $\lambda\neq 0,1$ orbital at a cell $x$ is connected via $P$ only to itself and exactly one partner orbital (with eigenvalue $1-\lambda$) either at cell $x-1$ or at $x+1$. On the other hand any orbital that corresponds to $\lambda=0$ is annihilated by $P$, while $\lambda=1$ orbitals are eigenvectors of $P$ with eigenvalue $1$. 
In this orbital basis, we straightforwardly obtain a generalized Wannier basis, i.e. an orthogonal basis of the image of $P$, with the property that each basis vector is also an eigenvector of $P\hat{x}P$.

Specifically, the basis consists of `monomers', i.e. all the $\lambda=1$ orbitals that exist, along with `dimers', each of which is a linear combination of a $\lambda$ orbital (for $\lambda\neq 0,1$), and its $(1-\lambda)$ partner orbital at a neighboring cell.
For example, if orbital $\ket{x,\lambda}$ corresponding to eigenvalue $\lambda\neq 0,1$ of $\p 0 x$ has a partner orbital $\ket{x+1,1-\lambda}$ at cell $x+1$, then $\p 1 x \ket{x,\lambda} = \sqrt{\lambda(1-\lambda)} \ket{x+1,1-\lambda}$.
Thus, we can define a state $\ket w$ which is a simultaneous eigenstate of $P$ and $P\hat{x}P$. Specifically,
\begin{align}
    \begin{split}
    \ket{w} & \coloneqq \sqrt \lambda \ket{x,\lambda} + \sqrt{1-\lambda} \ket{x+1,1-\lambda}, \\
    \text{is s.t. } P \ket w &= \ket w, \\
    \text{and } P\hat{x}P \ket w &= (x+1-\lambda)\ket w.
    \end{split}
\end{align}
The set of all such monomer and dimer states forms the set of generalized Wannier functions for $P$, since it spans the image of $P$, and consists of simultaneous eigenstates of $P$ and $P\hat{x}P$.

\subsection{Nearest Neighbor Projectors with Lattice Translational Invariance} \label{sec:NN with LTI}
We now return to the case of NN projectors with LTI. 
While the procedure for obtaining generalized Wannier functions also generates simultaneous eigenstates of $P$ and $P\hat{x}P$ operators for LTI NN projectors, the obtained basis may not be Wannier functions, since they may lack the property of forming a set that is invariant under any lattice translation operation.
We will now provide a modification that generates compact Wannier functions (which are also the maximally-localized Wannier functions).

First, we note that because of LTI, we have only two hopping matrices $P_1$ and $P_0$ [see \eref{eq:hopping_matrix_def}] that determine the projector, since $P(k) = P_0 + P_1 e^{ik} + P_1^\dagger e^{-ik}$. 
Additionally, many of the statements from the non-LTI case carry over.
Since $P_0^\dagger =P_0$ and $P_0 = P_0^2 + P_1 P_1^\dagger + P_1^\dagger P_1$, eigenvalues of $P_0$ are real and lie in $[0,1]$. 
As before, we diagonalize $P_0$ and obtain an orthonormal eigen-basis $\{ \ket{\lambda_i}\vert \ \lambda \text{ is an eigenvalue of }P_0\}$ for the orbital space $\h$, with $\lambda$s denoting $P_0$ eigenvalues and $i$ distinguishing degenerate states. 
Until specified otherwise, from this point onwards, by orbitals, we will mean vectors from this basis. 

Since $P_1 = P_1 P_0 + P_0 P_1$, we note that $\bra \mu P_1 \ket \lambda \neq 0$ for $\lambda\neq 0,1$ only if $\mu+\lambda=1$. Thus, $P$ connects orbitals $\lambda$ and $\mu$ at neighboring cells only if the sum of the $P_0$ eigenvalues they correspond to add to 1. We also note that if $P_0$ has an eigenvalue $\lambda\neq 0,1$, then it also must have an eigenvalue $1-\lambda$. So, if $n_\lambda$ denotes the multiplicity of eigenvalue $\lambda$, then for $\lambda \in (0,1)$, $n_\lambda\neq 0 \iff n_{1-\lambda} \neq 0$. Therefore, $P_1$ has a block structure similar to \eref{eq:block_structure_P1}. Dropping all the position indices from \eref{eq:block_structure_P1}, we can write it as 
\begin{widetext}
\begin{align}
    P_1 = \ 
    \begin{blockarray}{c|c|c|c|c|c|cc}
    \h_{\lambda} & \h_{1-\lambda} & \dotsc & \h_{\mu} & \h_{1-\mu} & \h_0 & \h_1  \\
    \begin{block}{(c|c|c|c|c|c|c)c}
    \cline{1-7}
    0&P_{(1-\lambda) \rightarrow \lambda}&0&0&0&0&0&\ \h_\lambda \\
    \cline{1-7}
    P_{\lambda\rightarrow (1-\lambda)}&0 &0&0&0&0& 0 &\ \h_{1-\lambda} \\
    \cline{1-7}
    0&0&\ddots&0&0&0&0&  \vdots \\
    \cline{1-7}
    0&0&0&0&P_{(1-\mu) \rightarrow \mu}&0&0&  \ \h_{\mu}\\
    \cline{1-7}
    0&0&0&P_{\mu \rightarrow (1-\mu)}&0&0&0&\ \h_{(1-\mu)}\\
    \cline{1-7}
    0&0&0&0&0& 0 &0&\ \h_0 \\
    \cline{1-7}
    0&0&0&0&0& 0 &0&\ \h_1 \\
    \end{block}
  \end{blockarray}, \label{eq:Block diagonal P1 TI}
\end{align}
\end{widetext}
with $\lambda,\dotsc,\mu$ denoting eigenvalues of $P_0$ that lie in $(0,0.5]$, and $\h_\lambda$ denoting the $\lambda$ eigenspace. 
$\lambda=1-\lambda$ when $\lambda=0.5$, so the two corresponding rows/columns collapse to one in this case.
Rows (and columns) corresponding to eigenvalues that do not exist for a particular example should be understood as being absent in the representation above.
Similarly to the procedure for the non-LTI projectors, we will obtain an SVD of the $P_1$ matrix by obtaining the SVDs of its various blocks. 

Let us first consider the blocks corresponding to $\lambda$ and $1-\lambda$ for $\lambda\in(0,0.5)$.
We start with an SVD of $P_{\lambda\rightarrow (1-\lambda)}$. 
If $\ket{ \lambda}$ and $\ket{ 1-\lambda}$ are two vectors corresponding to a non-zero singular value $\sigma$ so that $P_1 \ket{\lambda} = \sigma \ket{1-\lambda}$ and $P_1^\dagger\ket{1-\lambda}=\sigma\ket{\lambda}$, then $P_1 \ket{1-\lambda}=0$ and $P_1^\dagger \ket{\lambda} = 0$. 
[This follows from $P^2=P$ being an NN operator, so that $P_1^2=(P_1^\dagger)^2=0$.]
In position space, this means that $P$ maps the span of $\{ \ket{x,\lambda}, \ket{x+1,1-\lambda} \}$ onto itself. 
Thus, such vectors are `paired' with each other. 
As a consequence, $P_{(1-\lambda)\rightarrow \lambda}$ has possibly non-zero matrix elements only between vectors corresponding to zero singular values of $P_{\lambda\rightarrow (1-\lambda)}$.
An SVD of $P_{(1-\lambda)\rightarrow \lambda}$ can thus be implemented via rotations affecting only these subspaces, leaving the subspaces of the paired vectors untouched.
At the end of all these rotations, every $\lambda$ orbital at a cell is connected to one and only one $1-\lambda$ (`partner') orbital at a neighboring cell, and similarly for $1-\lambda$. 
We note again that all the non-zero singular values of both $P_{(1-\lambda)\rightarrow \lambda}$ and $P_{\lambda \rightarrow (1-\lambda)}$ are equal to $\sqrt{\lambda(1-\lambda)}$. This follows from the equation $P_0 = P_0^2 + P_1^\dagger P_1 + P_1 P_1^\dagger$.

Let us now consider the case of $P_{0.5\rightarrow 0.5}$. Suppose an SVD of this matrix is given by $P_{0.5\rightarrow 0.5} = \sum_{\sigma\neq 0}  \sigma \ket{\psi_\sigma}\bra{\phi_\sigma}$.
Since $P_1^2 = 0$, all the $\psi$ and $\phi$ vectors are orthogonal to each other. 
Thus, the set of all $\ket \psi$s and $\ket \phi$s form an orthonormal basis of $\h_{0.5}$. 
(This is because if there were to exist any $\ket \mu\in \h_{0.5}$ orthogonal to both $\ket \phi$s and $\ket \psi$s, then we would find that $P_1\ket{\mu}=P_1^\dagger \ket \mu = 0$, implying that $P \ket{x,\mu}=0.5 \ket{x,\mu}$, which is impossible.) 
In addition, $P_0 = P_0^2 + P_1^\dagger P_1 + P_1 P_1^\dagger$ implies that every singular value $\sigma=\sqrt{0.5 (1-0.5)}=0.5$. 
Thus, in this basis for $\h_{0.5}$, each orbital has a partner orbital, so that for every partner pair $\psi$ and $\phi$, $P$ maps the span of $\{ \ket{x,\phi}, \ket{x+1,\psi}\}$ onto itself.

It is now straightforward to construct Wannier functions using these orbitals. For each eigenvalue $\lambda\in (0,0.5]$ of $P_0$, every orbital $\ket{\lambda}$ is paired with exactly one orbital $\ket{1-\lambda}$. For every such pair, $\{ \sqrt{\lambda} \ket{x,\lambda} + \sqrt{1-\lambda} \ket{x+1,1-\lambda} \vert x\} $ forms one flavor of Wannier functions. Similarly, if an eigenvalue of $1$ exists, every corresponding eigenvector and all its lattice translates form a flavor of Wannier functions. It is straightforward to see that all these wavefunctions are eigenstates of $P\hat{x}P$ as well, and are thus the maximally-localized Wannier functions (MLWFs) for the span of the projector.

\subsection{Larger Hopping Range Projectors without Lattice Translational Invariance} \label{sec:supercell}
For lattice translationally invariant (LTI) strictly-local (SL) projectors with maximum hopping range $R>1$, as shown in the main text, we can always construct CWFs spanning its image, with each wavefunction having a maximum size of no more than $R+1$.
However, the technique is not applicable to projectors that are not LTI.
In order to obtain an orthogonal basis of compact wavefunction for such projectors with a maximum hopping range $R$, we first consider a supercell representation, so that $R$ cells in the original lattice are clubbed together to form one supercell. 
In the supercell representation, the SL projector is now an NN projector.
We can now apply the procedures from the previous subsections, and obtain a basis that spans the image of $P$. Reverting back to the original (i.e. primitive cell) lattice representation, this basis is then an orthogonal basis of compactly-supported wavefunctions that spans the image of $P$. 
These wave-functions then have a maximum spatial extent of no more than $2R$ cells. 

\section{Conclusion} \label{sec:conclusions}
Flat, or dispersion-less electronic bands in tight-binding models are spanned by compactly-supported Wannier-type functions, which in general can form a non-orthogonal set.
In line with the localization-topology correspondence, their existence is known to be incompatible with topological non-triviality of the band(s) they span. 
Wannier-type functions are nevertheless not true Wannier functions.  
In one-dimensional models, we answered the question of when one can form compact Wannier functions from compact Wannier-type functions.
We showed that the existence of compact Wannier functions is equivalent to the strict locality of the  band projector (or the single-particle Green's function). 
We provided a method for constructing compact Wannier functions corresponding to a strictly-local projector. 
We also showed that they are unique if they span a single band, and furthermore, compactly-supported Wannier-type functions are equivalent to compact Wannier functions when the latter exist.

For bands spanned by compact Wannier functions, we showed that maximally-localized Wannier functions are in general exponentially localized and not compactly supported, except when the band projector is nearest-neighbor hopping. 
In the latter case, we presented a procedure for obtaining maximally-localized Wannier functions (generalized Wannier functions) for projectors with (without) lattice translational invariance.
In the special case of nearest-neighbor hopping projectors, we showed how to obtain generalized Wannier functions even in the absence of lattice translational invariance.
We also presented a method for constructing all possible 1d models that have compact Wannier functions, which we expect will find applications in the construction of flat-band models with a rich variety of applications in single-particle and many-particle physics. 

A simple corollary of our work is that in higher dimensions, hybrid Wannier functions that are compactly supported along one direction can exist if and only if the band projector is strictly local along that direction.
Our methods do not directly apply to higher dimensions and it would be interesting to identify and prove conditions equivalent to the existence of compact Wannier functions in higher dimensions.
A lot of our results and discussion can also be rephrased in terms of the single-particle Green's function which suggest generalizations to interacting systems.
It would also be interesting to study compact Wannier functions and strictly local projectors in real continuum space instead of in tight-binding representations.

\section{Acknowledgements}
We thank Adrian Culver, Dominic Reiss, Steven Kivelson and Bartholomew Andrews for discussions, and Fenner Harper for collaboration on a related project. 
We acknowledge financial support from the University of California Laboratory Fees Research Program funded by the UC Office of the President (UCOP), grant number LFR-20-653926. 
P.S. acknowledges financial support from Bhaumik Graduate Fellowship (UCLA).

\appendix
\begin{widetext}
\section{Iterative Step Reduces the Range of the Projector}\label{sec:Projector range reduction appendix}
Here, we present details of the iterative procedure for obtaining compact Wannier functions for a given strictly local projector. We also present a procedure for obtaining maximally-localized Wannier functions (generalized Wannier functions) for nearest-neighbor hopping projectors with (without) lattice translational invariance.

In the main text, a procedure was presented for constructing compactly-supported Wannier functions (CWFs) for strictly-local projectors.
The procedure involves iteratively implementing unitary rotations, each of which reduces the maximum hopping range of the projector by $1$.
In this section, we will prove that the unitary rotations indeed reduce the range of the projector.

At the start of an intermediate step, let the projector have a maximum hopping range $r\geq 1$, and be denoted by $P^{(r)}$. Hence, in $k-$space, we can write $\p {} {(r)} = P_0 + \sum_{m=1}^r P_m e^{ikm} + P_m^\dagger e^{-ikm}$. We first compute an SVD of $P_r$:
\begin{align}
    P_r = \sum_{\sigma\neq 0} \sum_{i=1}^{n_\sigma} \sigma \ket{\psi_{\sigma, i}} \bra{\phi _{\sigma,i}}.
\end{align}
It follows from $P_r^2=0$ that all $\phi$'s and $\psi$'s are mutually orthogonal.
If the total number of these wavefunctions is less than $n$, the number of orbitals per cell, then one can straightforwardly obtain a set of wavefunction $\{ \ket \mu\}$ which completes the basis for $\h$ (i.e. the orbital space).

Now consider the unitary rotation 
\begin{align}
    \begin{split}
    U_r(k) &= \sum_{\sigma\neq 0}\sum_{i=1}^{n_\sigma} e^{-ik} \ket{\phi_{\sigma, i}} \bra{\phi_{\sigma, i}} \\
    & \ + \sum_{\sigma\neq 0}\sum_{i=1}^{n_\sigma} \ket{\psi_{\sigma,i}} \bra{\psi_{\sigma,i}} + \sum_{\mu} \ket{\mu}\bra{\mu}.
    \end{split} \label{eq:rotation one}
\end{align}
This is equivalent to first implementing an intra-cell rotation followed by a unit-cell redefinition.

In the new basis, the projector is represented by the matrix $U_r(k)^\dagger P(k) U_r(k)$. 
Clearly, all the matrix elements of $U_r(k)^\dagger P(k) U_r(k)$ are Laurent polynomials in $e^{ik}$.
We will now show that the degree of each polynomial in $U_r(k)^\dagger P(k) U_r(k)$ is less than $r$.

The calculation below is straightforward to carry out in the general case with multiple, possibly degenerate $\sigma$'s. 
For notational simplicity, we will go through the calculation only for the case with a single non-zero $\sigma$, with no degeneracy. 
This allows us to safely drop the subscripts $\sigma,i$ for $\psi$ and $\phi$ and proceed with a much simpler notation. 
We then have
\begin{align}
    P_r &= \sigma \ket{\phi}\bra{\psi}, \label{eq:SVD Pr single sigma} \\
    \text{and }U_r(k) &= e^{-ik}\ket{\phi} \bra{\phi} +  \ket{\psi} \bra{\psi} + \sum_{\mu} \ket \mu \bra \mu
\end{align}

It is straightforward to show that the coefficients of all powers greater than $r$ of $e^{ik}$ in $U_r^\dagger(k) P(k) U_r(k)$ are $0$. We will now show that the coefficient of $e^{ikr}$ is also zero. We have:
\begin{align}
    \begin{split}
    \text{Coefficient of }e^{ikr}\text{ in }U_r^\dagger (k) P(k) U_r(k)&= \text{Coefficient of }e^{ikr}\text{ in } \\
        & \left( e^{ik}\ket{\phi} \bra{\phi} +  \ket{\psi} \bra{\psi} + \sum_{\mu} \ket \mu \bra \mu \right) \left( P_0 + \sum_{m=1}^r P_m e^{ikm} + \sum_{m=1}^r P_m^\dagger e^{-ikm} \right)  \\
        & \times \left( e^{-ik}\ket{\phi} \bra{\phi} +  \ket{\psi} \bra{\psi} + \sum_{\mu} \ket \mu \bra \mu \right) \\
        &= \underbrace{\ket \phi \bra \phi P_r \ket \phi \bra \phi}_{t_1} + \underbrace{
        \Bigl(\ket \psi \bra \psi + \sum_{\mu} \ket \mu \bra \mu \Bigr) P_r \Bigl(\ket \psi \bra \psi + \sum_{\mu} \ket \mu \bra \mu \Bigr) }_{t_2}\\
        &+ \underbrace{
        \ket \phi \bra \phi P_{r-1} \Bigl( \ket \psi \bra \psi + \sum_\mu \ket \mu \bra \mu \Bigr)
        }_{t_3}\\
        &= 0. 
    \end{split}
\end{align}
This is because, $t_1$, $t_2$ and $t_3$ above are all zero.
That $t_1$ and $t_2$ vanish follows from \eref{eq:SVD Pr single sigma} and the mutual orthogonality of $\ket \psi$, $\ket \phi$ and $\ket \mu$'s. 
$t_3$ can be seen to vanish because $\bra \phi P_{r-1} \ket \mu= \bra \phi P_{r-1} \ket \psi= 0$.
To see this, note that for the case $r>1$, equating the coefficients of $e^{i(2r-1)k}$ on both sides of $P(k) = P(k)^2$ gives us
\begin{align}
    P_r P_{r-1} + P_r P_{r-1} &=0, \nonumber \\
    \text{and thus } \ket \psi \bra \phi P_{r-1} + P_{r-1} \ket \psi \bra {\phi} &= 0. \label{eq:intermediate step Pr Pr-1}
\end{align}
Let $\bra \alpha (\dotsc) \ket \beta$ denote left-multiplying by $\bra \alpha$ and right-multiplying by $\ket \beta$ both sides of equation $(\dotsc)$. Then, we have
\begin{align}
    \begin{split}
    \bra \phi \eref{eq:intermediate step Pr Pr-1} \ket \phi &\implies  \bra \phi P_{r-1} \ket \psi =0 \\
    \text{and } \bra \psi \eref{eq:intermediate step Pr Pr-1} \ket \mu &\implies  \bra \phi P_{r-1} \ket \mu =0.
    \end{split}
\end{align}
On the other hand, for $r=1$, we have 
\begin{align}
    P_0 P_1 + P_1 P_0 &= P_1 \nonumber  \\
    \therefore P_0 \ket \psi \bra \phi + \ket \psi \bra \phi P_0 &= \ket \psi \bra \phi,      \label{eq:intermediate step P1 P0}
\end{align}
so that
\begin{align}
    \begin{split}
    \bra \phi \eref{eq:intermediate step P1 P0} \ket \phi &\implies  \bra \phi P_0 \ket \psi =0 \\
    \text{and } \bra \psi \eref{eq:intermediate step P1 P0} \ket \mu &\implies  \bra \phi P_0 \ket \mu =0.
    \end{split}
\end{align}
Thus, in either case, $t_3=0$ and hence the largest positive power of $e^{ik}$ with a non-zero coefficient in $P_1(k)$ is less than $r$. 
It follows similarly that all powers of $e^{ik}$ that are $\leq (-r)$ are also zero, since $P(k)$ is a Hermitian matrix.

We have thus shown that the iterative step transforms an SL projector with a maximum hopping range $r$ to a projector with a maximum hopping range of at most $r-1$.

\section{Application of the Procedure to an Example Projector} \label{sec:example}
We now obtain compact Wannier functions for an example projector. 
Consider a strictly-local projector with a maximum hopping range of $2$, that in $k$-space is represented by 
\begin{align}
    P(k) = \begin{pmatrix}
            \frac{1}{3}+\frac{e^{-i k}}{6}+\frac{e^{i k}}{6} & -\frac{e^{-i k}}{6 \sqrt{2}}+\frac{e^{-2 i k}}{6 \sqrt{2}}-\frac{1}{3 \sqrt{2}} & \frac{e^{-i k}}{2 \sqrt{6}}+\frac{e^{-2 i k}}{2 \sqrt{6}} \\
            -\frac{e^{i k}}{6 \sqrt{2}}+\frac{e^{2 i k}}{6 \sqrt{2}}-\frac{1}{3 \sqrt{2}} & \frac{5}{12}-\frac{e^{-i k}}{6}-\frac{e^{i k}}{6} & \frac{1}{4 \sqrt{3}}-\frac{e^{-i k}}{2 \sqrt{3}} \\
            \frac{e^{i k}}{2 \sqrt{6}}+\frac{e^{2 i k}}{2 \sqrt{6}} & \frac{1}{4 \sqrt{3}}-\frac{e^{i k}}{2 \sqrt{3}} & \frac{1}{4} 
            \end{pmatrix} \label{eq:example SL projector range 2}
\end{align}
The coefficient of $e^{2ik}$, i.e. $P_2$ is 
\begin{align}
     P_2 &= \frac{1}{6\sqrt{2}}
            \begin{pmatrix}
                0 & 0 & 0 \\
                1 & 0 & 0 \\
                \sqrt{3} & 0 & 0 
            \end{pmatrix}
\end{align}
An SVD of $P_2 $ can be conveniently written as:
\begin{align}
    \begin{split}
    P_2 &= \frac{1}{3\sqrt{2}} \ket \psi \bra \phi \\
    \text{with } \ket{\psi} &= \frac{1}{2} \begin{pmatrix} 0 & 1 & \sqrt{3} \end{pmatrix}^T, \\
    \ket \phi &= \begin{pmatrix} 1 & 0 &  0 \end{pmatrix}^T
    \end{split}
\end{align}
To complete the basis for $\h$, we obtain $\ket \mu = \frac{1}{2} \begin{pmatrix} 0 & -\sqrt{3} & 1 \end{pmatrix} ^T$. The first rotation according to the procedure is then $U_2(k) = e^{ik} \ket \phi \bra \phi + \ket \psi \bra \psi + \ket \mu \bra \mu$ which in this case turns out to be $U_2(k) = \diag (e^{-ik}, 1, 1)$.
The projector after the rotation is represented by
\begin{align}
    \begin{split}
     P^{(1)} &= U_2 ^\dagger P U_2 \\
    &= \frac{1}{6} \begin{pmatrix}
 e^{-i k}+e^{i k}+2 & -\sqrt{2} e^{i k}+\frac{e^{-i k}}{\sqrt{2}}-\frac{1}{\sqrt{2}} & \sqrt{\frac{3}{2}}+\sqrt{\frac{3}{2}} e^{-i k} \\
 -\sqrt{2} e^{-i k}+\frac{e^{i k}}{\sqrt{2}}-\frac{1}{\sqrt{2}} & -e^{-i k}-e^{i k}+\frac{5}{2} & \frac{\sqrt{3}}{2}-\sqrt{3} e^{-i k} \\
 \sqrt{\frac{3}{2}}+\sqrt{\frac{3}{2}} e^{i k} & \frac{\sqrt{3}}{2}-\sqrt{3} e^{i k} & \frac{3}{2} \end{pmatrix}.
    \end{split}
\end{align}
As expected, all entries are Laurent polynomials with degrees less than or equal to 1. In the next step, we obtain an SVD of the $P_1$ corresponding to $P^{(1)}$:
\begin{align}
    \begin{split}
    P_1 &= \frac{1}{6} 
            \begin{pmatrix}    
                1 & -\sqrt{2} & 0 \\
                 \frac{1}{\sqrt{2}} & -1 & 0 \\
                 \sqrt{\frac{3}{2}} & -\sqrt{3} & 0 
             \end{pmatrix},\\
        \text{which has an SVD given by }P_1&= \frac{1}{2} \ket \psi \bra \phi \\
        \text{with }\ket \psi &= \frac{-1}{\sqrt{6}} 
                                \begin{pmatrix}
                                    \sqrt{2} & 1  & \sqrt{3}
                                \end{pmatrix} ^T, \\
        \ket \phi &= \frac{1}{\sqrt{3}}
                    \begin{pmatrix}
                    -1 & \sqrt{2} & 0 
                    \end{pmatrix}^T.
    \end{split}
\end{align}
The basis for $\h$ can be completed by adding the vector $\ket \mu = \begin{pmatrix}
 -\sqrt{2} & -1 & \sqrt{3} \end{pmatrix} ^T$ to the set $\{ \ket{\phi}, \ket{\psi}\}$. 
 The second rotation, $U_1 = e^{-ik}\ket{\phi} \bra \phi + \ket \psi \bra \psi + \ket \mu \bra \mu$ is then found to be
\begin{align}
    \begin{split}
    U_1 &= \frac{1}{3}\begin{pmatrix}
 2+e^{-i k} & \sqrt{2} \left(1-e^{-i k}\right) & 0 \\
 \sqrt{2} \left(1-e^{-i k}\right) & 1+2 e^{-i k} & 0 \\
 0 & 0 & 3 \end{pmatrix}.
    \end{split}
\end{align}
Implementing this rotation, we get 
\begin{align}
    \begin{split}
    \psup 1 \rightarrow \psup 0 &= U_1 ^\dagger \psup 1 U_1 \\
    &=\frac{1}{12}
    \begin{pmatrix}
        8 & -2 \sqrt{2} & 2 \sqrt{6} \\
        -2 \sqrt{2} & 1 & -\sqrt{3} \\
        2 \sqrt{6} & -\sqrt{3} & 3 \\
    \end{pmatrix}
    \end{split}    
\end{align}
The final step is diagonalizing $\psup 0$ with a unitary $U_0$:
\begin{align}
\begin{split}
    P^{(0)} \rightarrow U_0^\dagger P^{(0)} U_0 = &\diag (1,0,0)\\
    \text{with } U_0 &= 
    \begin{pmatrix}
        \sqrt{\frac{2}{3}} & -\sqrt{\frac{3}{11}} & \frac{1}{3} \\
        -\frac{1}{2 \sqrt{3}} & 0 & \frac{2 \sqrt{2}}{3} \\
        \frac{1}{2} & 2 \sqrt{\frac{2}{11}} & 0 \\
    \end{pmatrix}
    \end{split}
\end{align}
Thus, we have implemented a net unitary transformation $P\rightarrow U^\dagger P U = \diag (1,0,0)$, with $U = U_2 U_1 U_0$, given by 
\begin{align}
    U = \begin{pmatrix}
            \frac{e^{-2 i k} \left(1+e^{i k}\right)}{\sqrt{6}} & -\frac{e^{-2 i k} \left(1+2 e^{i k}\right)}{\sqrt{33}} & \frac{1}{3} e^{-2 i k} \left(-1+2 e^{i k}\right) \\
            \frac{e^{-i k} \left(-2+e^{i k}\right)}{2 \sqrt{3}} & -\sqrt{\frac{2}{33}} e^{-i k} \left(-1+e^{i k}\right) & \frac{1}{3} \sqrt{2} e^{-i k} \left(1+e^{i k}\right) \\
            \frac{1}{2} & 2 \sqrt{\frac{2}{11}} & 0 \\
        \end{pmatrix} \label{eq:unitary_for_example}
\end{align}
The family of Wannier function $\{ \ket w_x \vert x\in \z \}$ corresponding to $P$ is obtained by simply taking the inverse Fourier transform of the first column of $U$. 
Clearly, $\ket w_x$ (up to lattice translation and phase re-definitions) is
\begin{align}
    \begin{split}
    \ket{w}_x &= \ket{x} \otimes \begin{pmatrix}
        \frac{1}{\sqrt{6}} \\ 0 \\ 0
    \end{pmatrix} 
    + \ket{x+1} \otimes \begin{pmatrix}
        \frac{1}{\sqrt{6}} \\ \frac{-1}{\sqrt{3}} \\ 0
    \end{pmatrix} \\
    + & \ \ket{x+2} \otimes \begin{pmatrix}
        0 \\ \frac{1}{2\sqrt{3}} \\ \frac{1}{2}
    \end{pmatrix}
    \end{split}
\end{align}
As a corollary, we also obtain Wannier functions for $(I-P)$, which in this case are the inverse Fourier transforms of the second and third columns of \eref{eq:unitary_for_example}.

\end{widetext}

\end{document}